\documentclass{PoS}

\title{B meson excitations with chirally improved light quarks 
\footnote{For the BGR Collaboration.}}

\ShortTitle{B meson excitations with CI light quarks}

\author{\speaker{T.~Burch}, D.~Chakrabarti, C.~Hagen, T.~Maurer, 
  A.~Sch\"afer\\
  Institut f\"ur Theoretische Physik, Universit\"at Regensburg, 
  D-93040 Regensburg, Germany\\
  E-mail: \email{tommy.burch@physik.uni-regensburg.de}}

\author{C.~B.~Lang and M.~Limmer\\
  Institut f\"ur Physik, FB Theoretische Physik, Universit\"at Graz, 
  A-8010 Graz, Austria}

\abstract{
  We present our latest results for the excitations of static-light mesons 
  on both quenched and unquenched lattices, where the light quarks are 
  simulated using the chirally improved (CI) lattice Dirac operator.
}

\FullConference{The XXV International Symposium on Lattice Field Theory\\
		 30 July -- 4 August 2007\\
		 Regensburg, Germany}

\begin{document}

\section{Light-quark propagator estimation}

To enhance the signals of our static-light correlators, we use an improved 
estimate of the light-quark propagator from any point within half of the 
lattice to any point in the other half. 
This so-called ``domain decomposition improvement'' was outlined and tested 
in Ref.\ \cite{Burch:2006mb} and amounts to a variant of the ``maximal 
variance reduction'' approach \cite{Michael:1998sg}. 
We present the basics of the method here.

Decomposing the lattice into two distinct domains, the full Dirac matrix can 
be written in terms of submatrices 
\begin{equation}
  M = \left(
    \begin{array}{cc}
      M_{11} & M_{12} \\
      M_{21} & M_{22}
    \end{array} \right) \, ,
\end{equation}
where $M_{11}$ and $M_{22}$ connect sites within a region and 
$M_{12}$ and $M_{21}$ connect sites from the different regions. 
We can also write the propagator in this form: 
\begin{equation}
  M^{-1} = P = \left(
    \begin{array}{cc}
      P_{11} & P_{12} \\
      P_{21} & P_{22}
    \end{array} \right) \, .
\end{equation}
The propagator between regions 1 and 2 is then estimated using $N$ random 
sources ($\chi^n$, $n=1,..,N$): 
\begin{eqnarray}
  P_{12} &=& - M^{-1}_{11} M_{12} P_{22} \nonumber \\
  &\approx& - M^{-1}_{11} M_{12} \frac1N \chi_2^{n} \chi_2^{n\dagger} P \nonumber \\
  &\approx& -\frac1N \left( M^{-1}_{11} M_{12} \chi_2^{n} \right) \left( \gamma_5 P \gamma_5 \chi^{n}_2 \right)^{\dagger} 
  \equiv -\frac1N \psi^{n}_1 \phi^{n\dagger}_2 \; .
\end{eqnarray}
Note that no sources are needed in region 1 and those in region 2 should 
reach region 1 with one application of $M$. 
Since $M$ is usually a sparse matrix, this greatly reduces the number of 
lattice sites which the random sources cover.
\footnote{
  Even for the case of a non-sparse lattice Dirac operator (Overlap or 
  low-mode-subtracted), one may separate (or ``dilute'') the sources into, 
  for example, those close to the boundary and those further away: e.g., 
  $\chi^{n}_2 = \chi^{n}_{2,\;t=t_{bound}} + \chi^{n}_{2,\;t>t_{bound}}$.
}

In the following, we use the chirally improved (CI) lattice Dirac operator 
\cite{CI_action} for $M$.

\section{Static-light correlators}

Using different ``wavefunctions'' for the light-quark source and sink, we 
construct the following matrix of correlators: 
\begin{eqnarray}
  C_{ij}(t) &=& \left\langle 0 \left| (\bar Q \, O_j \, q)_t \;
      (\bar q \, \bar O_i \, Q)_0 \right| 0 \right\rangle \nonumber \\
  &=& \left\langle \sum_x \mbox{Tr} \left[
      \frac{1+\gamma_4}{2}\prod_{k=0}^{t-1} U_4^\dagger(x+k\hat{4}) \,
      O_j P_{x+t\hat{4},x} \bar O_i
    \right] \right\rangle \; ,
\end{eqnarray}
where $x$ is in one domain and $x+t\hat{4}$ is in the other.

\begin{table}
\begin{center}
\begin{tabular}{lcc} \hline
oper. & $J^P$ & $O(\Gamma,\vec D)$ \\ \hline
$S$ & $0^-,1^-$ & $\gamma_5$ \\
$P_-$ & $0^+,1^+$ & $\sum_i \gamma_i D_i$ \\
$P_+$ & $1^+,2^+$ & $\gamma_1 D_1 - \gamma_2 D_2$ \\
$D_\pm$ & $1^-,2^-,3^-$ & $\gamma_5(D_1^2 - D_2^2)$ \\ \hline
\end{tabular}
\caption{
  Static-light meson operators.
}
\label{sl_ops}
\end{center}
\end{table}

We use bilinears of the form: 
\begin{equation}
  \bar Q \, O_j \, q = \bar Q \; O(\Gamma,\vec D) \, 
  \left( \vec D^2 \right)^{l_j} \, S_J(\kappa,N_{sm,j}) \; q \; ,
\end{equation}
where $S_J$ is a gauge-covariant (Jacobi) smearing function and we apply 
$l_j=0$, 1, or 2 Laplacians. We also include the local source to obtain a 
$4 \times 4$ correlator matrix for each set of quantum numbers, determined 
by $O(\Gamma,\vec D)$ (see Table \ref{sl_ops}). 
The parameters used for smearing the light-quark sources and the details of 
the configurations we use \cite{Lang:2005jz} are given in Table \ref{params}.

\begin{table}
\begin{center}
\begin{tabular}{cccccc} \hline
$N_S^3 \times N_T$ & $a$ (fm) & $M_{\pi,\mbox{sea}}$ & link smear & $N_{conf}$ & 
$\left( ^{l_1}_{l_3}\;^{l_2}_{l_4} \; , \; ^{N_{sm,1}}_{N_{sm,3}}\;^{N_{sm,2}}_{N_{sm,4}} \; , \; \kappa \right)$ \\ \hline \vspace*{6pt}
$12^3 \times 24$ & 0.20 & $\infty$ & Hyp & 200 & $\left( ^0_1\;^0_2 \; , \; ^0_{12}\;^{8}_{16} \; , \; 0.2 \right)$ \\ \vspace*{6pt}
$16^3 \times 32$ & 0.15 & $\infty$ & Hyp & 100 & $\left( ^0_1\;^0_2 \; , \; ^0_{18}\;^{12}_{24} \; , \; 0.2 \right)$ \\ 
$16^3 \times 32$ & 0.16 & 450 MeV & Stout & 40 & $\left( ^0_1\;^0_2 \; , \; ^0_{18}\;^{12}_{24} \; , \; 0.2 \right)$ \\ \hline
\end{tabular}
\caption{
  Parameters for the configurations and quark source smearings.
}
\label{params}
\end{center}
\end{table}

\section{Mass splittings}

Once we have our correlator matrices, we apply the variational method 
\cite{Var_Meth} and solve the generalized eigenvalue problem 
\begin{equation}
  \sum_j C_{ij}(t) \; \nu_j^k = \lambda^k(t,t_0) 
  \sum_j C_{ij}(t_0) \; \nu_j^k \; .
\end{equation}
The eigenvalues behave as 
\begin{equation}
  \lambda^k(t,t_0) \propto e^{-t \, M_k} 
  \left[1+O(e^{-t \, \Delta M_k})\right] \; ,
\end{equation}
where $\Delta M_k$ is the difference to the state closest in mass to $M_k$. 
To help stabilize the matrix diagonalization, we check that our correlator 
matrices are real and symmetric (within errors) and then symmetrize them 
before solving the eigenvalue problem (via Cholesky decomposition). 
Although in principle one should work at the largest possible value of $t_0$, 
we find a negligible $t_0$-dependence in the eigenvalues and effective masses 
(and their jackknife errors) over the region where it is still possible to 
invert $C(t_0)$. So we present results where $t_0/a=1$.

\begin{figure}
\begin{center}
\includegraphics*[width=3.7cm]{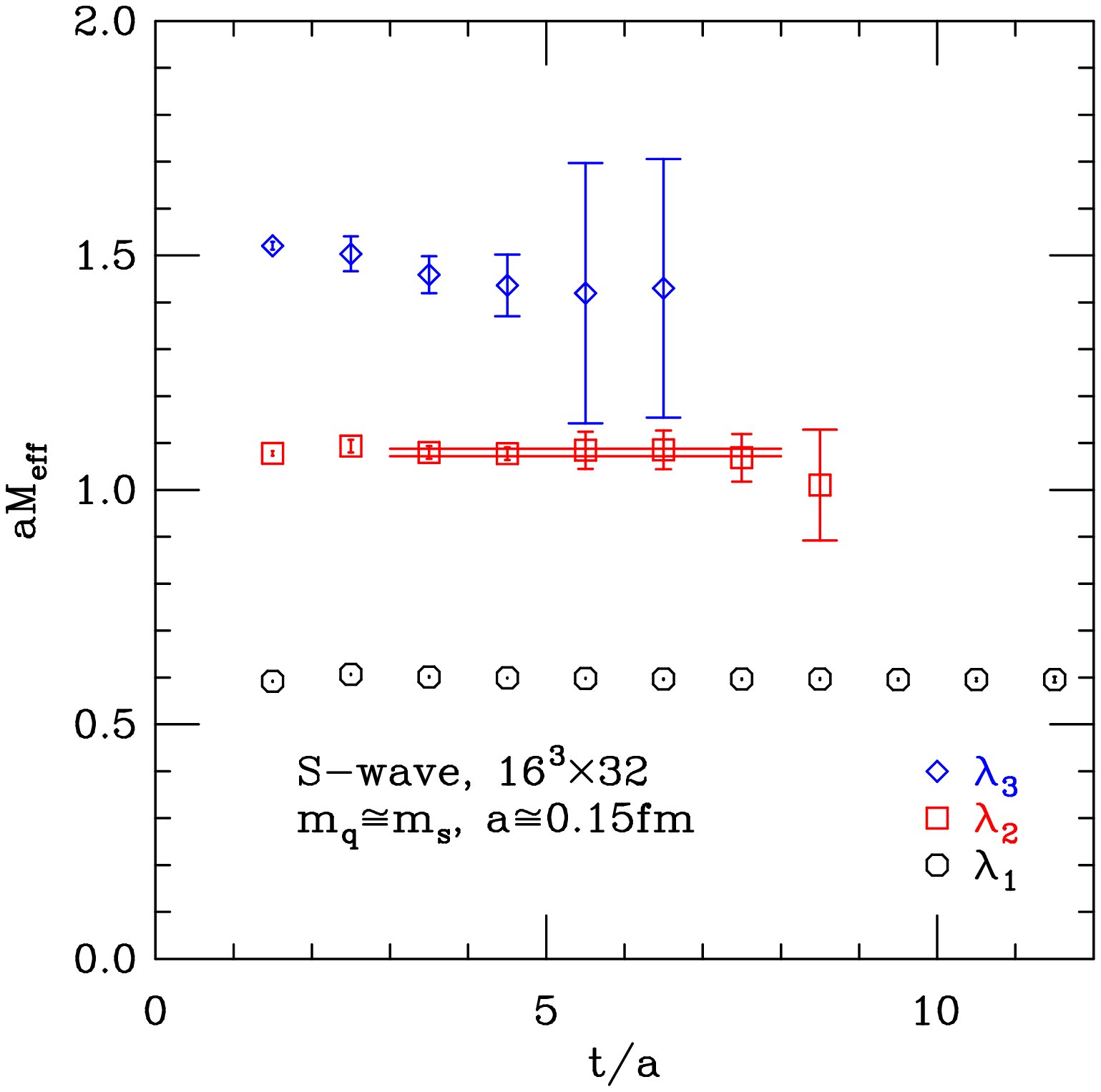}
\includegraphics*[width=3.7cm]{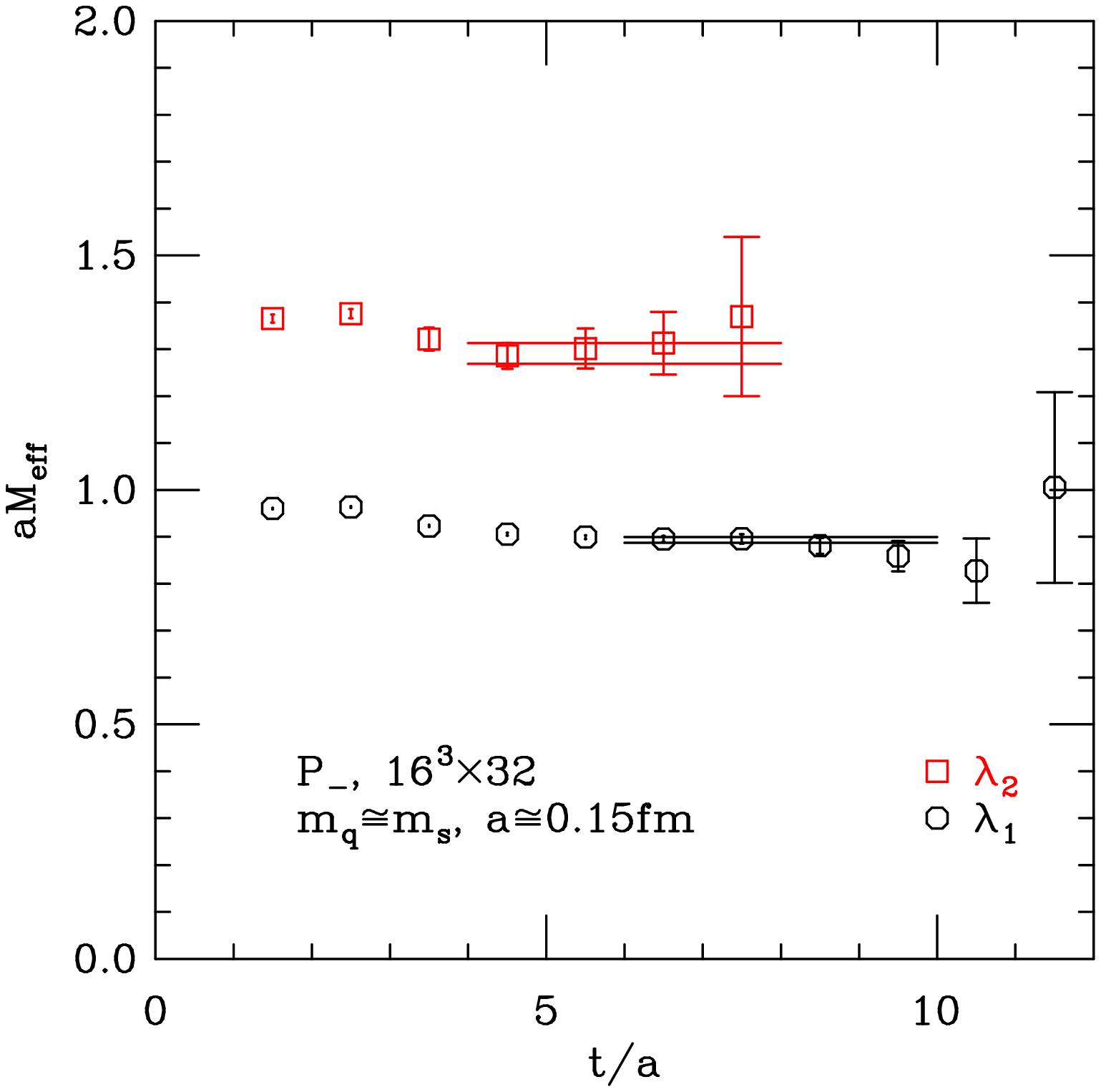}
\includegraphics*[width=3.7cm]{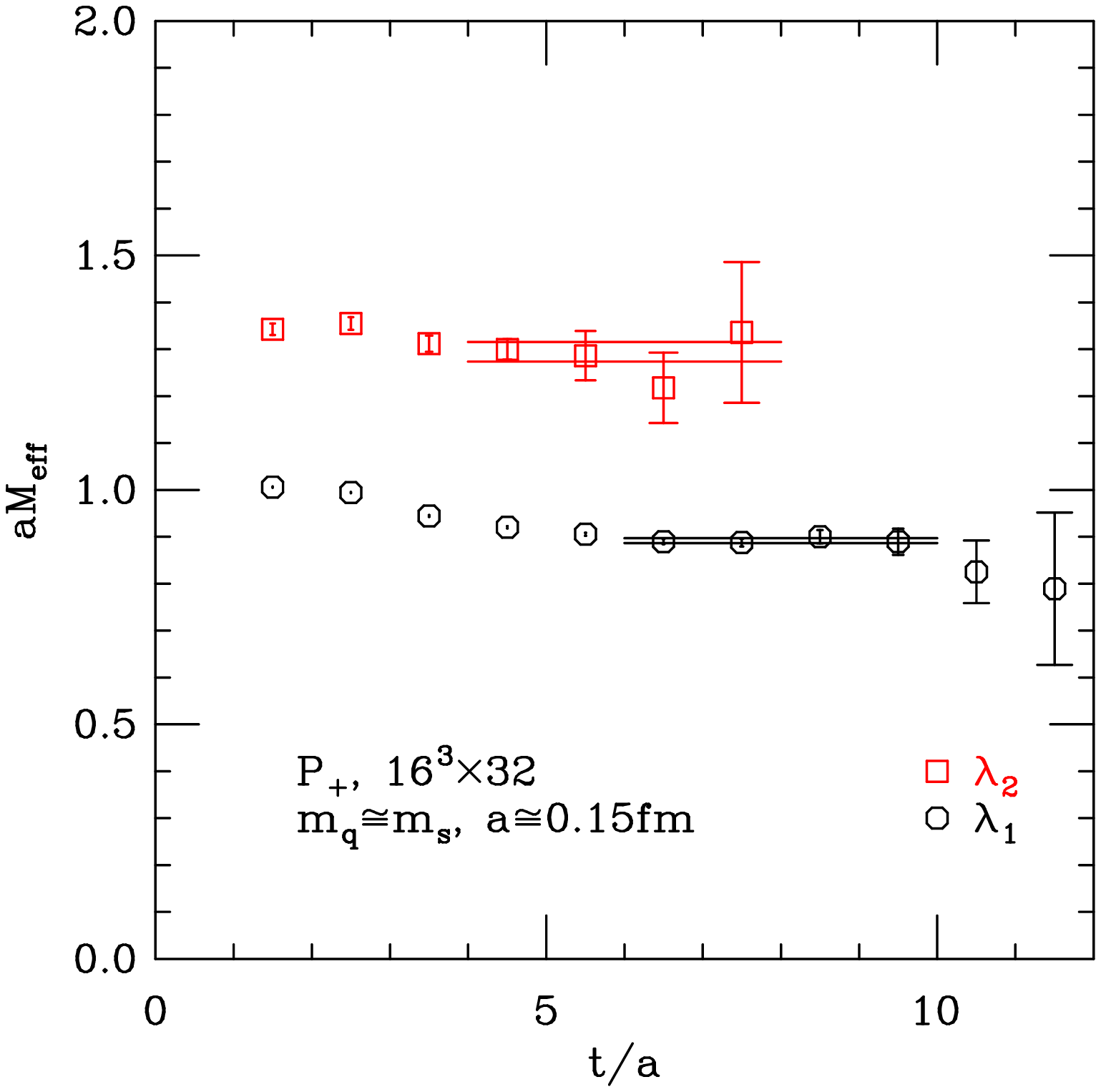}
\includegraphics*[width=3.7cm]{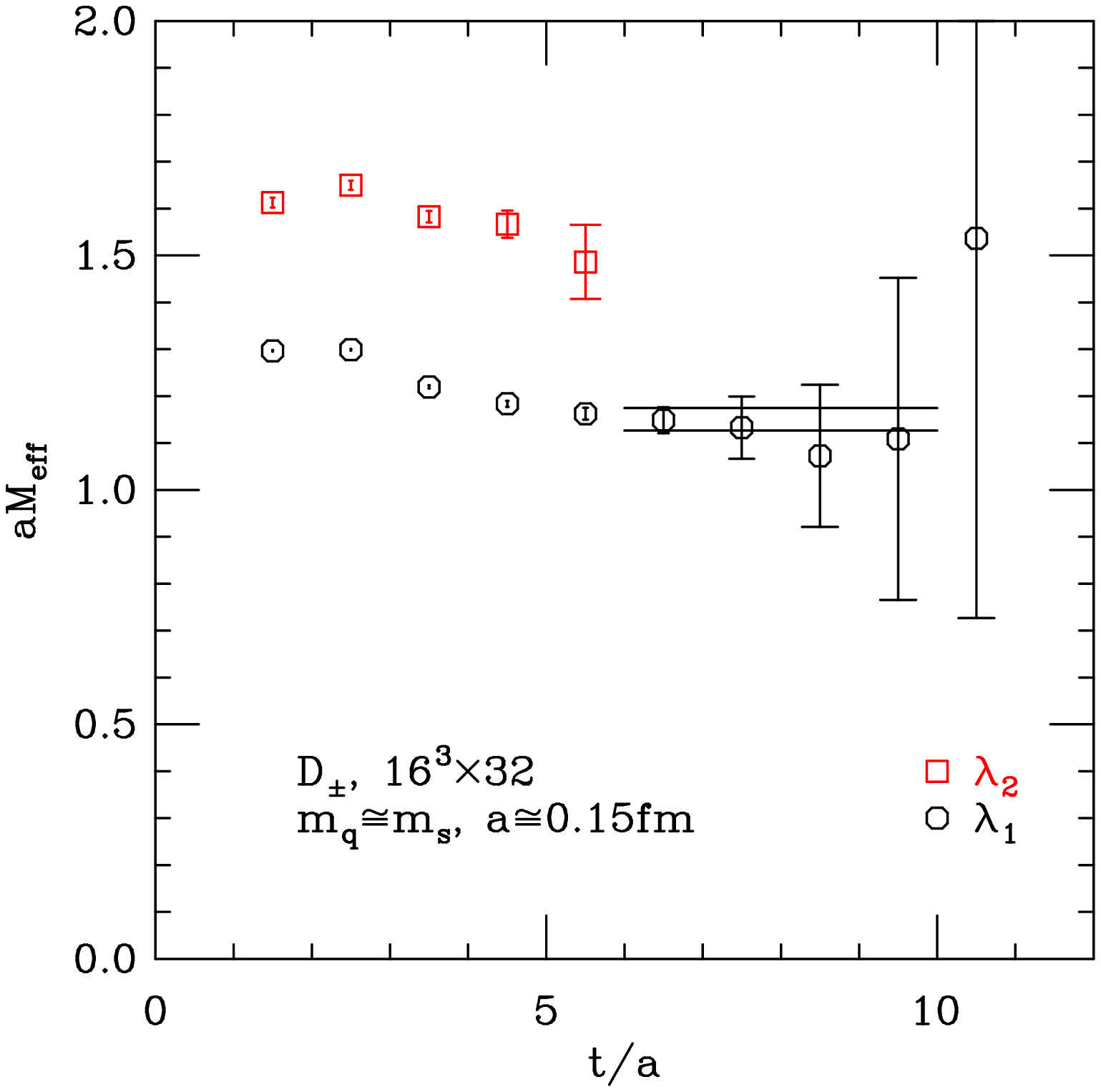}
\end{center}
\caption{
Effective masses for the static-light mesons on the quenched configurations. 
$am_q=0.08$, $a^{-1} \approx 1330$ MeV, $L \approx 2.4$ fm. 
The horizontal lines represent $M\pm\sigma_M$ fit values for the 
corresponding time ranges.
}
\label{effmass_que}
\end{figure}

\begin{figure}
\begin{center}
\includegraphics*[width=3.75cm]{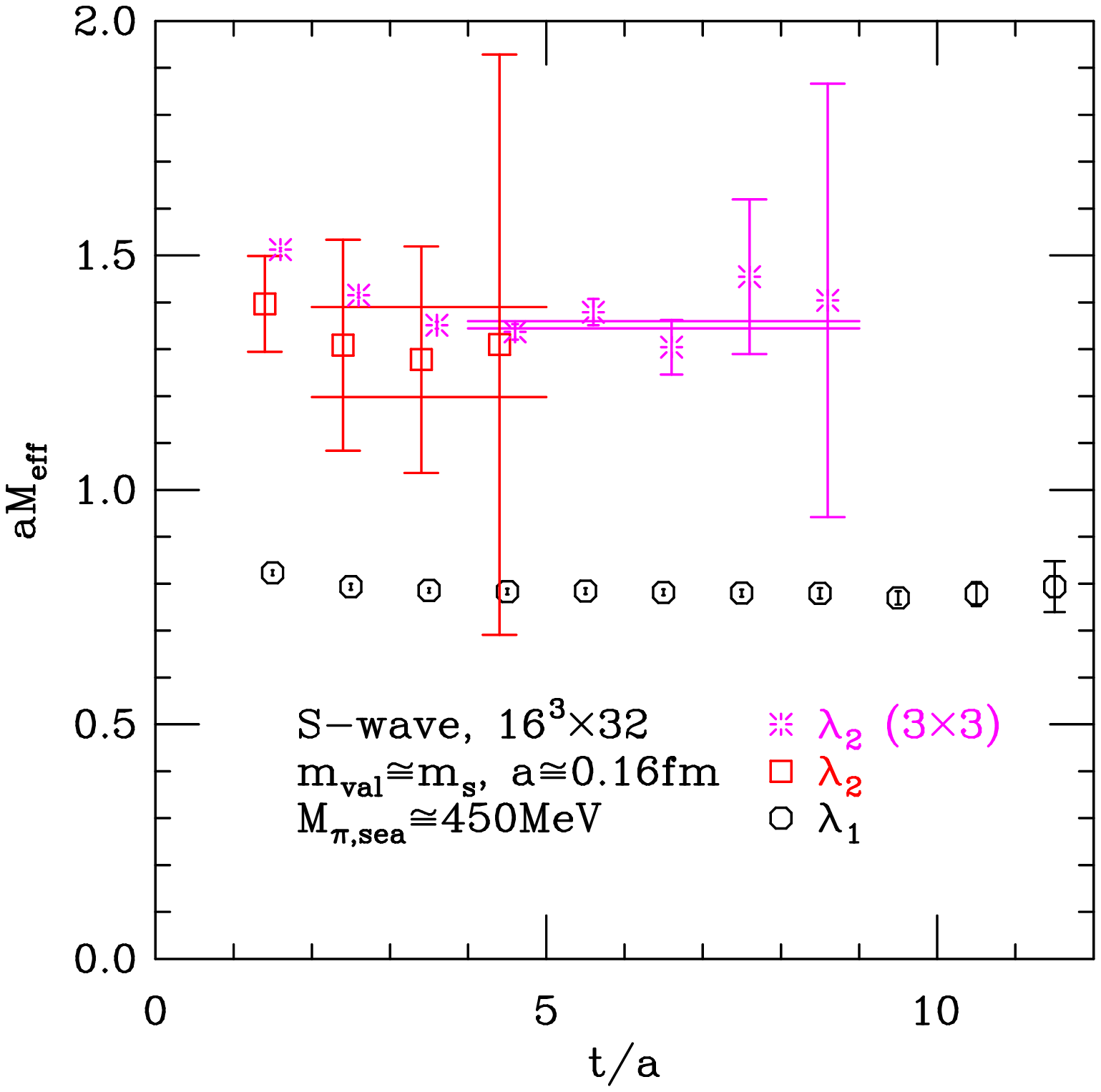}
\includegraphics*[width=3.6cm]{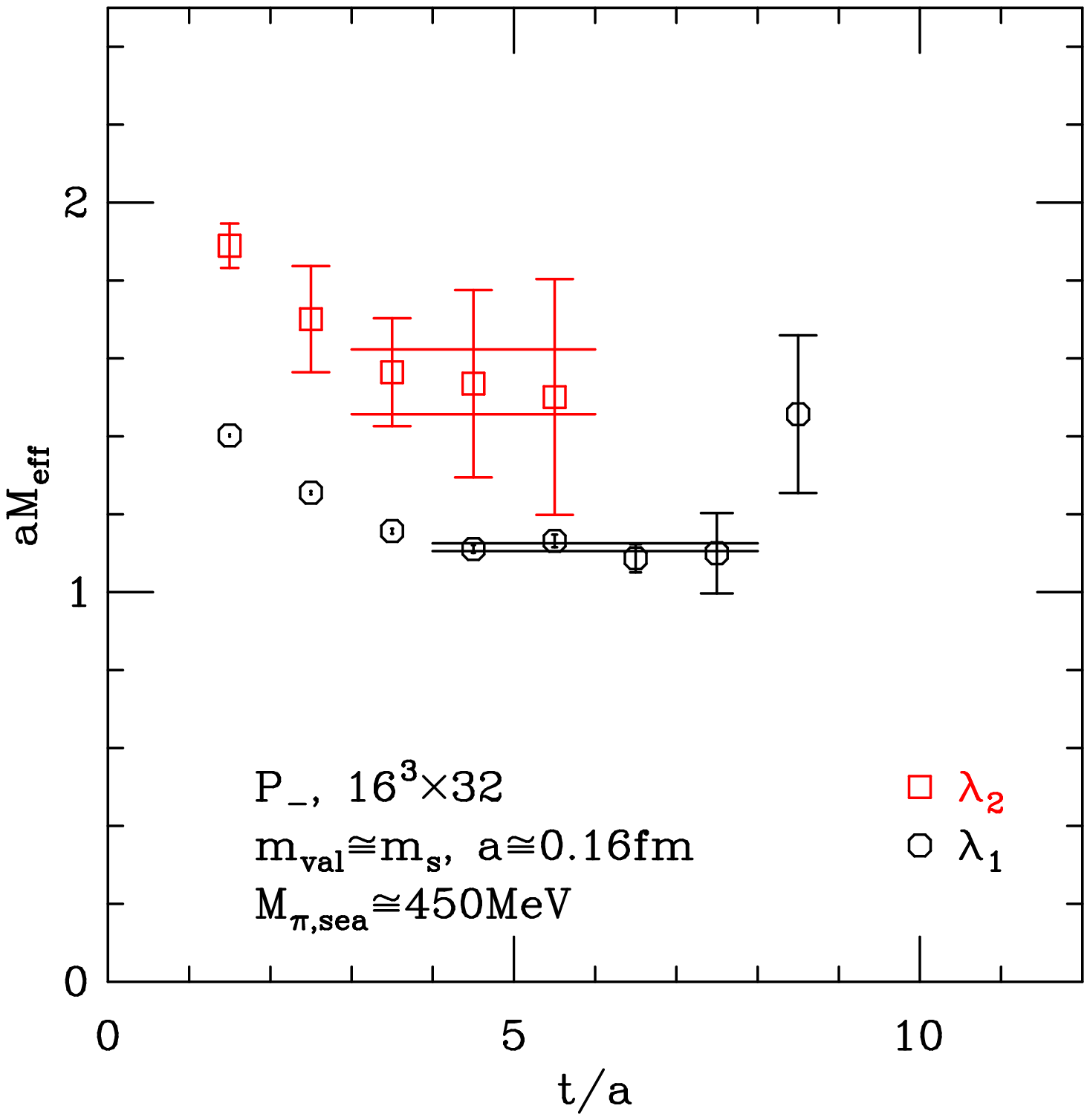}
\includegraphics*[width=3.6cm]{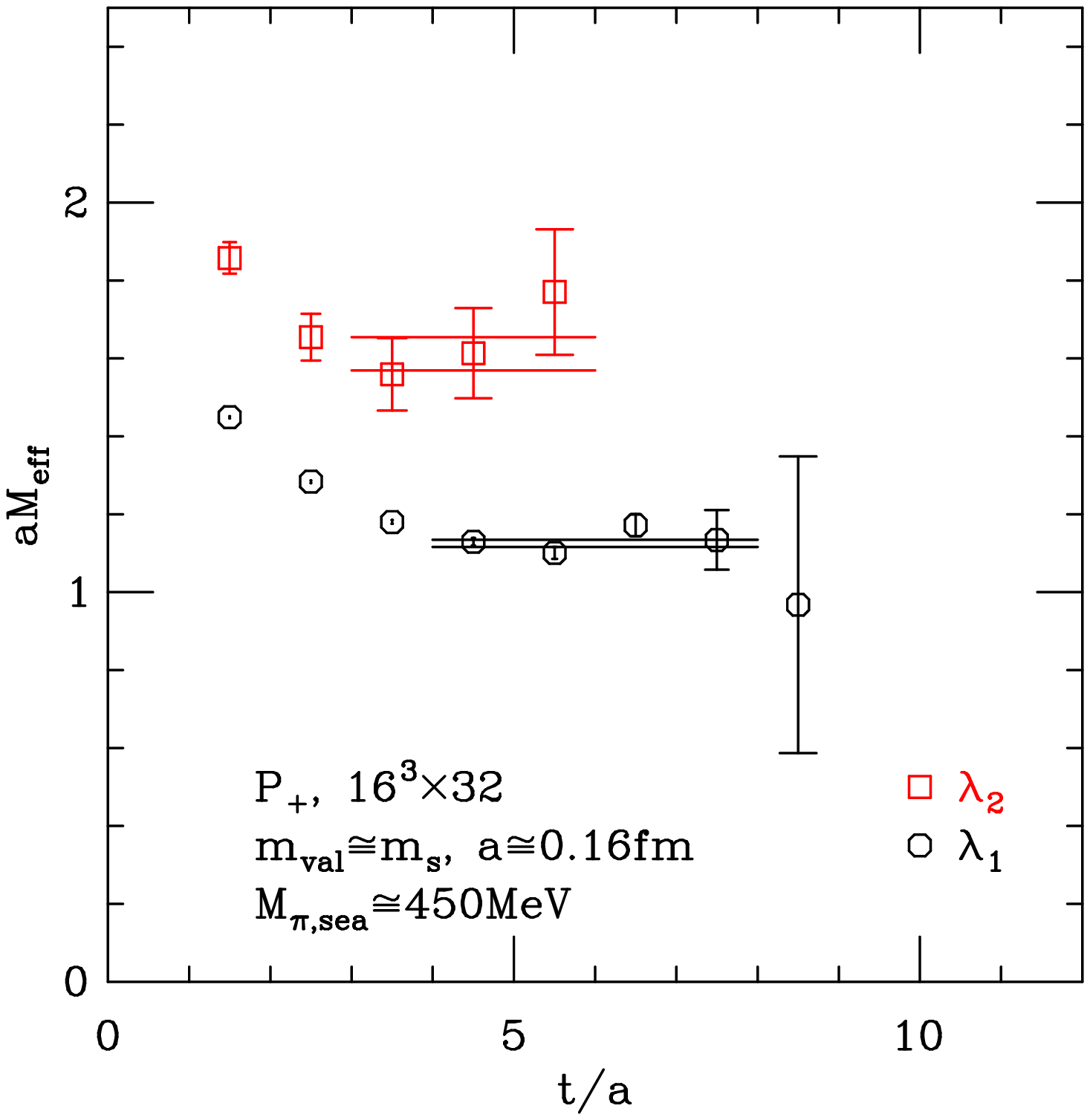}
\includegraphics*[width=3.6cm]{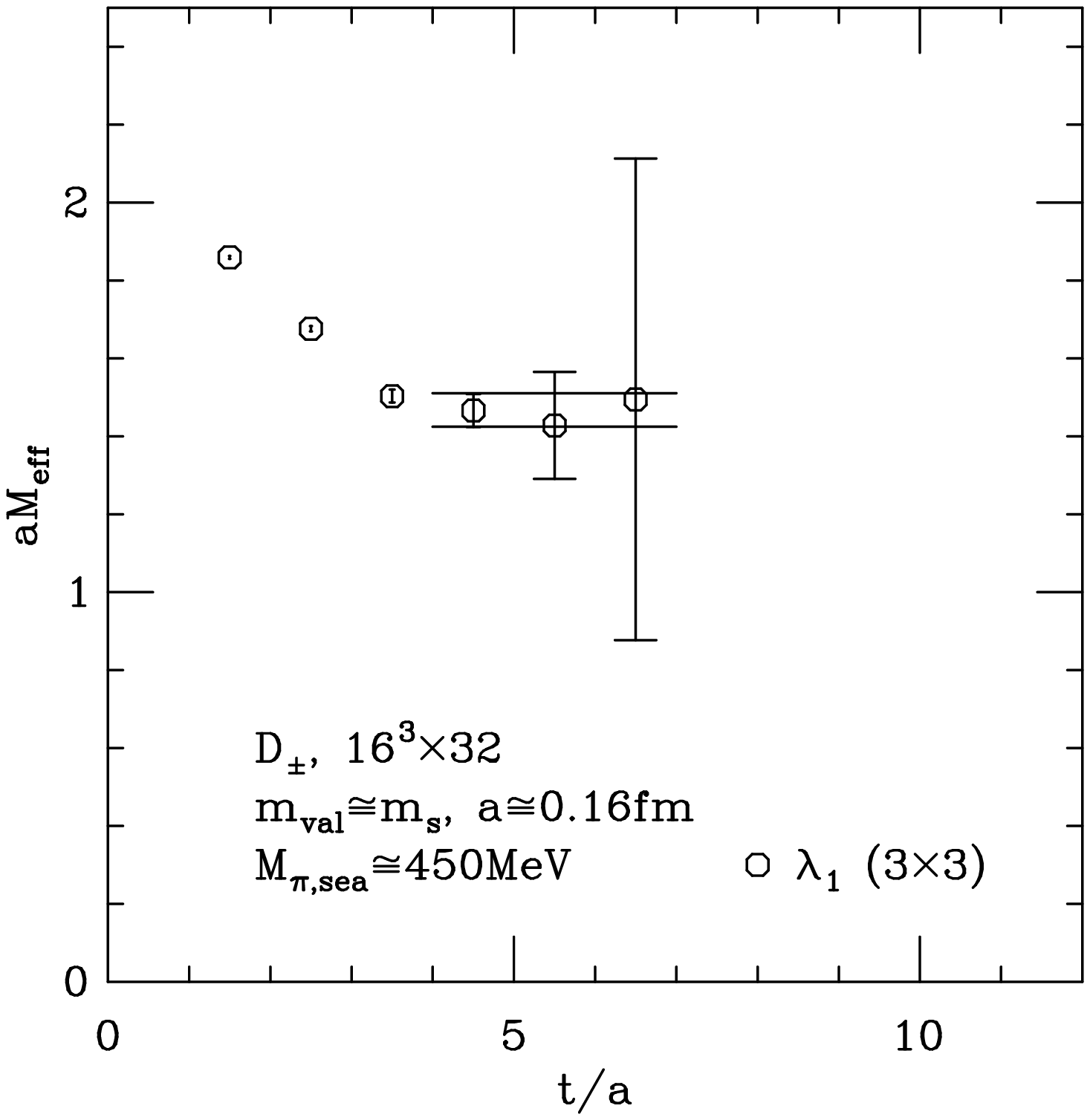}
\end{center}
\caption{
Effective masses for the static-light mesons on the dynamical configurations. 
$m_q\approx m_s$, $a^{-1} \approx 1230$ MeV, 
$M_{\pi,\mbox{sea}} \approx 450$ MeV, $L \approx 2.5$ fm. 
The horizontal lines represent $M\pm\sigma_M$ fit values for the 
corresponding time ranges.
}
\label{effmass_dyn}
\end{figure}

In Fig.~\ref{effmass_que} we show some of the effective masses which result 
from the $16^3 \times 32$ quenched configurations. Figure \ref{effmass_dyn} 
displays the effective masses from the dynamical configurations. 
In each figure appear the $S$-, $P_-$-, $P_+$-, and $D_\pm$-waves from 
left to right, respectively. 
The horizontal lines represent $M\pm\sigma_M$ fit values for the 
corresponding time ranges.

\begin{figure}
\begin{center}
\includegraphics*[width=10cm]{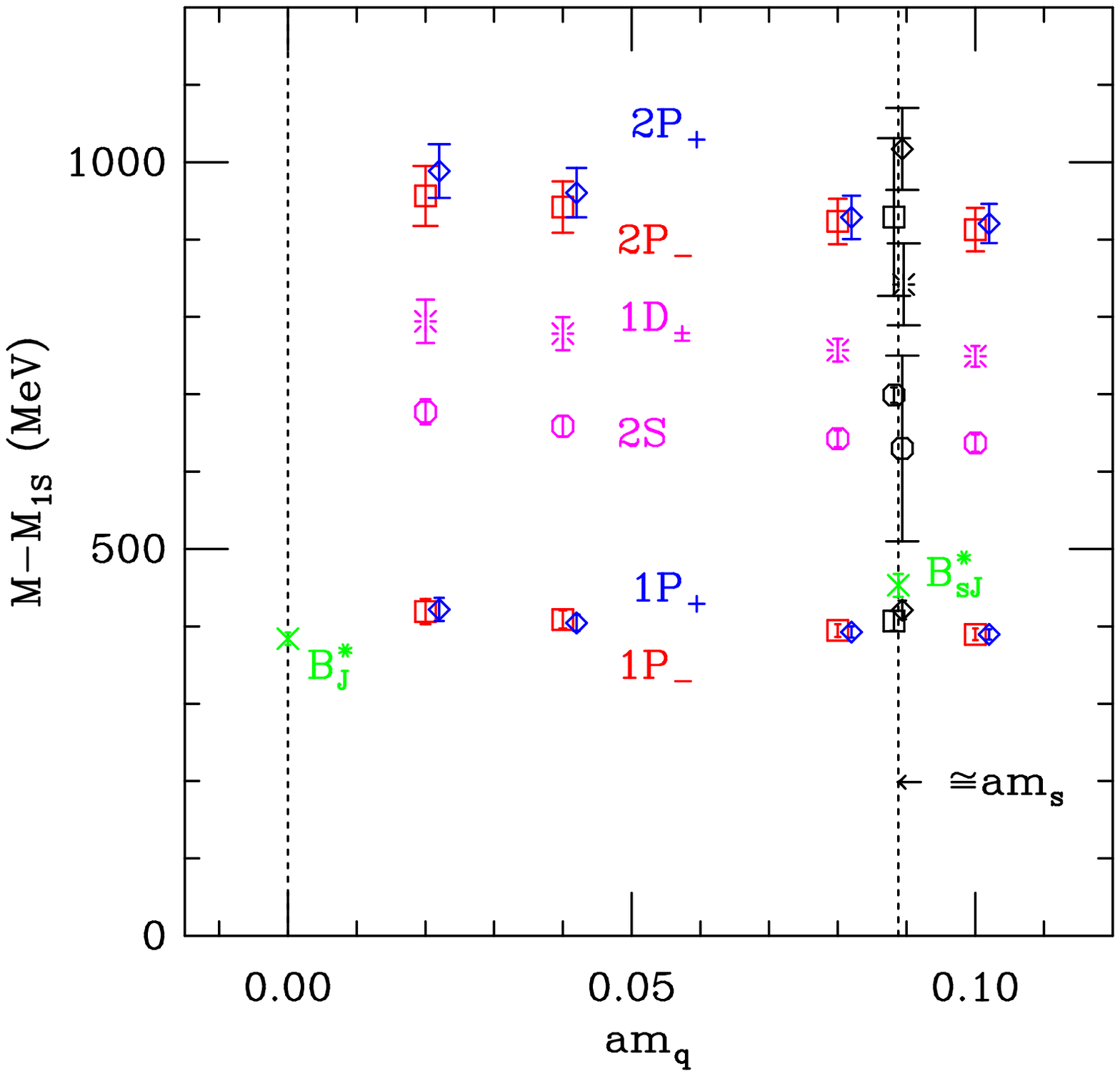}
\end{center}
\caption{
Mass splittings ($M-M_{1S}$) in MeV as a function of the quark mass for 
the $16^3\times32$ lattices. 
The black points at $m_s$ represent the results from the dynamical lattice. 
The squares are for $P_-$ results, diamonds are for $P_+$, circles are for 
$2S$, and bursts are for $1D_\pm$. 
The green $\times$'s are the experimental masses for $B_{(s)J}^*$ 
(differenced from $3M_{B_{(s)}^*}/4+M_{B_{(s)}}/4$) from the PDG \cite{PDG}.
}
\label{chiral_extrap}
\end{figure}

Figure \ref{chiral_extrap} displays the mass differences $M-M_{1S}$ on the 
$16^3 \times 32$ quenched lattice as a function of the light-quark mass. 
The green crosses at $m_q=m_s$ and $m_q=0$ are the experimental masses for 
$B_{sJ}^*$ and $B_{J}^*$ (differenced from $3M_{B_{(s)}^*}/4+M_{B_{(s)}}/4$), 
respectively, taken from the PDG \cite{PDG}. 
The black symbols at $m_q=m_s$ are the results from the dynamical lattice. 
The symbols have the same meaning as the colored ones for the quenched case.

\section{Discussion}

In Table \ref{Bs_splits} we report our $B_s$ meson mass splittings in 
physical units for the three lattices considered. 
One can see here that the $1P-1S$ splitting is too small when compared with 
experiment (as opposed to the $m_q \to 0$ case, where it appears too 
high; see Fig.~\ref{chiral_extrap}). 
Also, with statistical errors of $\sim 10$ MeV, the $1P_+-1P_-$ splitting is 
not resolved, except on the coarser quenched lattice, where it is 
$\sim 40$ MeV. 
We plan to study this further with a finer quenched lattice and higher 
statistics for the dynamical lattice. 
We would also like to try to include $1/m_Q^{}$ effects by interpolating 
between our results ($m_Q^{}=\infty$) and the experimental results for $D_s$ 
mesons (see \cite{Green:2003zz}).

It will also be interesting to watch the $2S-1S$ splitting 
(holding thus far $\sim 650 - 700$ MeV for $m_q \approx m_s$; see also 
\cite{Burch:2006mb}) 
as we proceed to higher statistics and finer lattice spacing.

It is important to keep in mind the possibly additional systematic error 
introduced by setting the scale of our lattices (we use $r_0^{}=0.5$ fm). 
A smaller value ($r_0^{} \approx 0.45-0.5$ fm; see, e.g., 
\cite{McNeile_plenary07}) would enhance our mass splittings.

\begin{table}
\begin{center}
\begin{tabular}{lcccc} \hline
state & \multicolumn{4}{c}{$M-M_{1S}$ (MeV)} \\
 & {\small $M_{\pi,\mbox{sea}}=\infty$} & {\small $M_{\pi,\mbox{sea}}=\infty$} & {\small $M_{\pi,\mbox{sea}} \approx 450$ MeV} & {\small PDG \cite{PDG}} \\
 & {\small $a \approx 0.20$ fm} & {\small $a \approx 0.15$ fm} & {\small $a \approx 0.16$ fm} & \\ \hline
$2S$ & 684(14) & 640(11) & 699(10)($^{+40}_{-180}$) & - \\ \hline
$1P_-$ & 385(7) & 393(8) & 407(13) & 453(15) \\
$2P_-$ & 995(20) & 918(29) & 930(100) & - \\ \hline
$1P_+$ & 422(4)($^{+0}_{-7}$) & 391(7) & 421(12) & 453(15) \\
$2P_+$ & 967(17) & 925(27) & 1017(53) & - \\ \hline
$1D_\pm$ & 730(12) & 755(14) & 842(53) & - \\
$2D_\pm$ & 1210(30) & - & - & - \\ \hline
\end{tabular}
\caption{
  Our static-light meson mass splittings at $m_q=m_s$. 
  Numbers in the first set of parentheses are statistical errors. 
  The second set (if present) represent the changes in the error bounds when 
  shifting to another seemingly good fit range (for the dynamical $2S$ state, 
  this is accompanied by an increase in the basis from the first 3 operators 
  to the full 4).
}
\label{Bs_splits}
\end{center}
\end{table}

\acknowledgments

We would like to thank Christof Gattringer for helpful discussions. 
Simulations were performed at the LRZ in Munich. 
This work is supported by GSI. 
The work of D.C.\ is supported by the Alexander von Humboldt Foundation.

\end{document}